\documentclass[aps,prl,reprint,twocolumn,superscriptaddress,floatfix,nofootinbib,longbibliography]{revtex4-1}
\usepackage{graphicx,amsmath,amsfonts,amssymb,amsthm,xr}
\usepackage{epsfig,amsmath,amssymb,color,dsfont,upgreek,physics}
\usepackage{mathrsfs}
\usepackage{mathtools}
\usepackage{bbold}
\usepackage{comment}
\usepackage{float}

\usepackage[bookmarks=true,colorlinks,linkcolor=OrangeRed,urlcolor=NavyBlue,citecolor=RoyalBlue]{hyperref}
\usepackage[dvipsnames]{xcolor}
\usepackage{orcidlink}

\definecolor{mygold}{rgb}{0.93,0.69,0.13}

\definecolor{mypurple}{rgb}{0.49,0.18,0.56}

\begin{document}
\title{Quantum Many-Body Scarring in $2+1$D Gauge Theories with Dynamical Matter}
\author{Jesse Osborne${}^{\orcidlink{0000-0003-0415-0690}}$}
\affiliation{School of Mathematics and Physics, The University of Queensland, St.~Lucia, QLD 4072, Australia}

\author{Ian P.~McCulloch${}^{\orcidlink{0000-0002-8983-6327}}$}
\affiliation{Department of Physics, National Tsing Hua University, Hsinchu 30013, Taiwan}

\author{Jad C.~Halimeh${}^{\orcidlink{0000-0002-0659-7990}}$}
\email{jad.halimeh@physik.lmu.de}
\affiliation{Department of Physics and Arnold Sommerfeld Center for Theoretical Physics (ASC), Ludwig-Maximilians-Universit\"at M\"unchen, Theresienstra\ss e 37, D-80333 M\"unchen, Germany}
\affiliation{Munich Center for Quantum Science and Technology (MCQST), Schellingstra\ss e 4, D-80799 M\"unchen, Germany}
\affiliation{Dahlem Center for Complex Quantum Systems, Freie Universit\"at Berlin, 14195 Berlin, Germany}

\begin{abstract}
Quantum many-body scarring (QMBS) has emerged as an intriguing paradigm of weak ergodicity breaking in nonintegrable quantum many-body models, particularly lattice gauge theories (LGTs) in $1+1$ spacetime dimensions. However, an open question is whether QMBS exists in higher-dimensional LGTs with dynamical matter. Given that nonergodic dynamics in $d{=}1$ spatial dimension tend to vanish in $d{>}1$, it is important to probe this question. Using matrix product state techniques for both finite and infinite systems, we show that QMBS occurs in the $2{+}1$D $\mathrm{U}(1)$ quantum link model (QLM), as evidenced in persistent coherent oscillations in local observables, a marked slowdown in the growth of the bipartite entanglement entropy, and revivals in the fidelity. Interestingly, we see that QMBS is more robust when the matter degrees of freedom are bosonic rather than fermionic. Our results further shed light on the intimate connection between gauge invariance and QMBS, and highlight the persistence of scarring in higher spatial dimensions. Our findings can be tested in near-term analog and digital quantum simulators, and we demonstrate their accessibility on a recently proposed cold-atom analog quantum simulator.
\end{abstract}

\date{\today} 
\maketitle

\textbf{\textit{Introduction.---}}QMBS is an exciting paradigm of ergodicity breaking in isolated quantum many-body models that are expected to thermalize \cite{Bernien2017,Moudgalya2018,Zhao2020,Jepsen2021,Serbyn2020,Moudgalya_review,Chandran_review}. Despite being ergodic, certain models host special nonthermal \textit{scar} eigenstates that are roughly equally spaced in energy over the whole spectrum \cite{Turner2018,Schecter2019}, and exhibit anomalously low bipartite entanglement entropy \cite{BernevigEnt,lin2018exact}. Upon preparing an initial state with a high overlap with these scar eigenstates, the system avoids thermalization and its dynamics exhibits long-lived coherent oscillations lasting well beyond all relevant timescales, with the time-evolved wave function undergoing persistent periodic revivals \cite{Turner2018,wenwei18TDVPscar}. QMBS is of great importance in investigations of the Eigenstate Thermalization Hypothesis (ETH) \cite{Deutsch1991,Srednicki1994,Rigol_review,Deutsch_review}, as it facilitates violations of the latter through novel mechanisms based on spectrum-generating algebras \cite{MotrunichTowers,MoudgalyaHubbard,Dea2020,Pakrouski2020} and nonthermal-eigenstate embedding \cite{ShiraishiMori}. QMBS has also been the subject of various experiments in both analog and digital quantum simulators \cite{Bernien2017,Bluvstein2021,Bluvstein2022quantum,Su2022,Zhang2022Many-body,Dong2023Disorder}.

QMBS also has an intimate connection to lattice gauges theories (LGTs) \cite{Surace2020,Iadecola2020quantum,aramthottil2022scar,biswas2022scars,Desaules2022weak,Desaules2022prominent,Halimeh2022robust}, which are interacting quantum many-body models hosting gauge symmetries that enforce an intrinsic relation between the local distribution of matter and the allowed corresponding configurations of the gauge fields \cite{Weinberg_book,Peskin2016}. Indeed, the quantum Ising-like model realized in a Rydberg setup that produced the first instance of QMBS \cite{Bernien2017} can be effectively described by the PXP model, which Surace \textit{et al.}~\cite{Surace2020} have shown to map exactly onto the $1{+}1$D spin-$1/2$ $\mathrm{U}(1)$ QLM \cite{Chandrasekharan1997,Wiese_review}. The spin-$1/2$ $\mathrm{U}(1)$ QLM is a formulation of the Schwinger model in which the gauge and electric fields are represented by spin-$1/2$ operators. This truncation has facilitated experimental feasibility in the realization of large-scale quantum simulators of the spin-$1/2$ $\mathrm{U}(1)$ QLM \cite{Yang2020,Zhou2022,Su2022,Zhang2023observation}, with proposed algorithms for digital platforms \cite{Barros2024meroncluster}, while still capturing salient features of the Schwinger model such as Coleman's phase transition \cite{Coleman1976}. Given the ongoing strong drive of realizing quantum simulators of LGTs \cite{Dalmonte_review,Pasquans_review,Zohar_review,Alexeev_review,aidelsburger2021cold,Zohar_NewReview,klco2021standard,Bauer_review,dimeglio2023quantum,halimeh2023coldatom,cheng2024emergent}, it is important to fully understand the connection between QMBS and LGTs both from a fundamental point of view and also to facilitate experimental investigations. 

In particular, it is interesting to see how QMBS behaves in $2{+}1$D LGTs. Indeed, it is known that nonergodic features in $1{+}1$D interacting models tend to vanish in higher spatial dimensions $d{>}1$, with many-body localization \cite{Basko2006,Gornyi2005,Nandkishore_review,Abanin_review} being a prime example. How the \textit{weak} ergodicity-breaking mechanism of QMBS fares for $d{=}2$ is a question that has recently received some attention in the context of the PXP model \cite{Michailidis2020,Lin2020quantum} and helix states in the XXZ model \cite{Jepsen2021}, in addition to other quasi-$2{+}1$D systems \cite{Zhang2022Many-body,Dong2023Disorder}. Nevertheless, scarred models in $d{>}1$ are far and few in between compared to their counterparts in $d{=}1$ \cite{Serbyn2020,Moudgalya_review,Chandran_review}. This further motivates adding to the collection of such models by investigating LGTs with dynamical matter in higher spatial dimensions.

Indeed, QMBS in LGTs in $d{>}1$ has been studied \cite{Banerjee2021,biswas2022scars,Sau2024}, but only in the case without dynamical matter, where connections to high-energy phenomena are not as straightforward. In trying to see whether QMBS will persist in the quantum-field-theory limit of gauge theories, such as in quantum electrodynamics, the inclusion of dynamical matter is important \cite{Desaules2022weak,Desaules2022prominent}.

In this Letter, we consider the $2{+1}$D spin-$1/2$ $\mathrm{U}(1)$ QLM with dynamical matter on a square lattice, and show that QMBS persists for special far-from-equilibrium quenches. We also showcase how these QMBS regimes can be detected in near-term analog quantum simulators of LGTs in $d{=}2$ spatial dimensions.

\textbf{\textit{Model.---}}We consider the $\mathrm{U}(1)$ QLM on a square lattice described by the Hamiltonian~\cite{Chandrasekharan1997,Wiese_review}
\begin{align}\label{eq:qlm}
    \hat{H} {=} \sum_\mathbf{r} \bigg [ &{-}\frac{\kappa}{2} \left( \hat{\phi}_{\mathbf{r}}^\dagger \hat{s}^+_{\mathbf{r},\mathbf{e}_x} \hat{\phi}_{\mathbf{r}+\mathbf{e}_x}{+} \alpha (-1)^{r_y} \hat{\phi}_{\mathbf{r}}^\dagger \hat{s}^+_{\mathbf{r},\mathbf{e}_y} \hat{\phi}_{\mathbf{r}+\mathbf{e}_y} {+} \text{H.c.} \right) \nonumber\\
    & {+} m (-1)^{r_x+r_y} \hat{\phi}_{\mathbf{r}}^\dagger \hat{\phi}_{\mathbf{r}}{-}\alpha^2 J \left( \hat{U}_{\Box_\mathbf{r}}{+} \hat{U}_{\Box_\mathbf{r}}^\dagger \right) \bigg].
\end{align}
The matter degrees of freedom with mass $m$ are represented by the ladder operators $\hat{\phi}^{(\dagger)}_\mathbf{r}$ at site $\mathbf{r}$.
These can either be fermionic or hard-core bosonic, and we will consider both in this Letter.
The gauge (electric) field is represented by the spin-$1/2$ operator $\hat{s}^{+(z)}_{\mathbf{r},\mathbf{e}_a}$, where the subscript denotes the bond between sites $\mathbf{r}$ and $\mathbf{r}{+}\mathbf{e}_a$, where $\mathbf{e}_a$ is a unit vector ($a{=}x,y$). 
The magnetic interactions of the gauge fields, whose strength is proportional to \(J\), are represented by the plaquette operators \(\hat{U}_{\Box_\mathbf{r}} = \hat{s}^+_{\mathbf{r},\mathbf{e}_x} \hat{s}^+_{\mathbf{r}+\mathbf{e}_x,\mathbf{e}_y} \hat{s}^-_{\mathbf{r}+\mathbf{e}_y,\mathbf{e}_x} \hat{s}^-_{\mathbf{r},\mathbf{e}_y}\).
The coefficient \(\alpha\) is used to tune the ratio of the coupling along the \(y\) and \(x\) axes, such that the couplings are equally strong at \(\alpha = 1\), and at \(\alpha = 0\), only the coupling along the \(x\) axis remains, and the system behaves like a collection of uncoupled \(1+1\)D QLMs.
Thus, by using \(0 < \alpha < 1\), we can interpolate between the \(2+1\)D square lattice (\(\alpha = 1\)) and the \(1+1\)D model (\(\alpha = 0\)).

\begin{figure}[t]
    \includegraphics{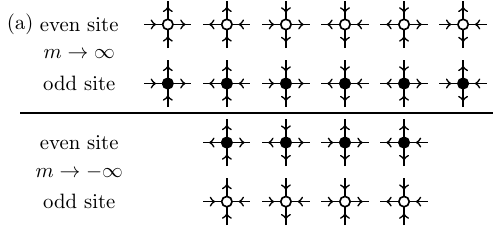}
    \includegraphics{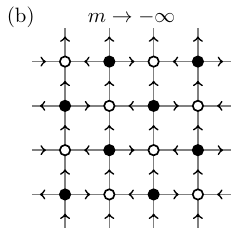}\quad~
    \includegraphics{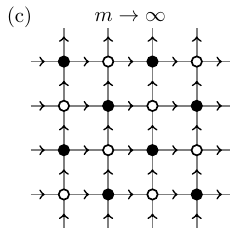}
    \caption{(a) The gauge-invariant configurations for the gauge sites surrounding each matter site in the spin-1/2 \(\mathrm{U}(1)\) quantum link model~\eqref{eq:qlm}, according to Gauss’s law~\eqref{eq:gauss}.
    The site parity refers to the parity of the sum of the \(x\) and \(y\) components of the index: \(r_x+r_y\).
    Arrows pointing right or up on gauge sites represent eigenstates of \(\hat{s}^z\) with eigenvalue \(+1/2\), while arrows pointing left or down represent the eigenvalue \(-1/2\).
    (b) A charge-proliferated ground state at \(m \rightarrow -\infty\).
    (c) A vacuum ground state at \(m \rightarrow \infty\).}
    \label{fig:1}
\end{figure}

The \(\mathrm{U}(1)\) gauge symmetry of this model is generated by the operators
\begin{equation}\label{eq:gauss}
    \hat{G}_\mathbf{r} = \hat{\phi}_\mathbf{r}^\dagger \hat{\phi}_\mathbf{r} {-} \frac{1 {-} (-1)^{r_x{+}r_y}}{2} {-} \sum_{a=x,y} \big[ \hat{s}^z_{\mathbf{r},\mathbf{e}_a} {-} \hat{s}^z_{\mathbf{r}-\mathbf{e}_a,\mathbf{e}_a} \big],
\end{equation}
which act as a discrete analog of Gauss’s law.
We work in the physical sector of states, which are eigenstates of \(\hat{G}_\mathbf{r}\) for each \(\mathbf{r}\) with eigenvalue zero.
Given a configuration of the matter sites, this becomes a constraint on the allowed configuration on the gauge sites: the allowed configurations surrounding occupied and unoccupied matter sites are shown in Fig.~\ref{fig:1}(a).

\textbf{\textit{Scarring dynamics.---}}Here we consider the dynamics following the global quench of an initial charge-proliferated state \(\ket{\psi(t=0)}\), being the gauge-invariant ground state at \(m/\kappa \rightarrow -\infty\), using the gauge site configuration shown in Fig.~\ref{fig:1}(b) (which is chosen such that at \(\alpha = 0\), the decoupled chains are in the physical gauge sector of the $1+1$D model).
We quench to a finite value of the mass \(m\), which is known to lead to scarred dynamics in the $1+1$D model~\cite{Su2022,Hudomal2022,Daniel2023}, in particular, we quench to \(m = 0.84\kappa\), which is the regime considered in the experiment of Ref.~\cite{Su2022}, while setting $J=0$.
We use numerical time-evolution simulations to obtain these dynamics based on matrix product state (MPS) techniques~\cite{Uli_review,Paeckel_review,mptoolkit}.
Specifically, we use the time-dependent variational principle (TDVP) algorithm~\cite{Haegeman2016}, using single-site updates with adaptive bond dimension expansion.
Simulations were performed using a cylindrical lattice geometry, with a circumference of \(L_y = 4\) matter sites: we perform calculations using states which are explicitly translation invariant along the \(x\)-axis, as well as states with a finite length \(L_x = 16\) in order to calculate the fidelity with the initial state, using open boundaries at the ends of the cylinder.

\begin{figure}[t]
    \includegraphics{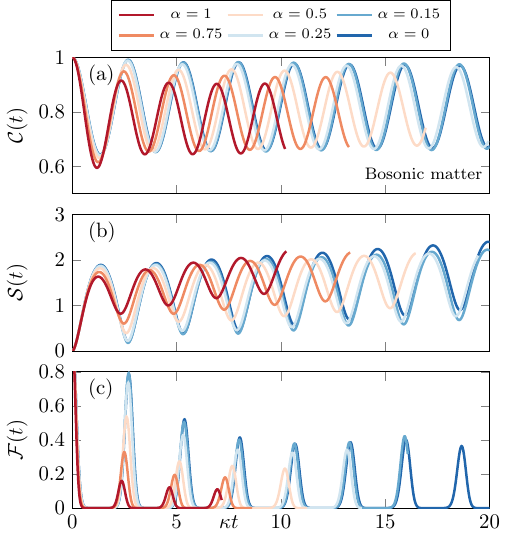}
    \caption{A numerical time-evolution simulation of the quench of the charge-proliferated state shown in Fig.~\ref{fig:1}(b) to \(m = 0.84 \kappa\), using matter with bosonic statistics on a \(L_y = 4\) cylinder, as the coupling ratio \(\alpha\) tuned from 1 to 0.
    (a) The chiral condensate \(\mathcal{C}(t)\)~\eqref{eq:chiral-condensate} and (b) entanglement entropy across a circumferential slice \(\mathcal{S}(t)\), calculated for an infinite-length cylinder.
    (c) The fidelity \(\mathcal{F}(t)\), calculated for a finite cylinder of dimension \(4 \times 16\) with open boundaries in the \(x\) direction.}
    \label{fig:quench-b-cp}
\end{figure}

\begin{figure}[t]
    \includegraphics{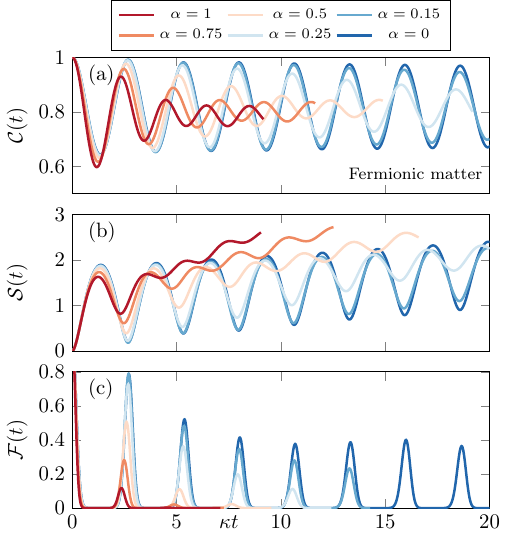}
    \caption{A numerical time-evolution simulation of the quench of the charge-proliferated state shown in Fig.~\ref{fig:1}(b) to \(m = 0.84 \kappa\), using matter with fermionic statistics on a \(L_y = 4\) cylinder, as the coupling ratio \(\alpha\) tuned from 1 to 0.
    (a) The chiral condensate \(\mathcal{C}(t)\)~\eqref{eq:chiral-condensate} and (b) entanglement entropy across a circumferential slice \(\mathcal{S}(t)\), calculated for an infinite-length cylinder.
    (c) The fidelity \(\mathcal{F}(t)\), calculated for a finite cylinder of dimension \(4 \times 16\) with open boundaries in the \(x\) direction.}
    \label{fig:quench-f-cp}
\end{figure}

The quench simulation results are displayed in Figures~\ref{fig:quench-b-cp} and \ref{fig:quench-f-cp}, for bosonic and fermionic matter statistics respectively, as the coupling ratio \(\alpha\) is tuned from one to zero.
In panels (a), we show the expectation value of the chiral condensate \(\mathcal{C}(t) = \mel{\psi(t)}{\hat{\mathcal{C}}}{\psi(t)}\),
\begin{equation}\label{eq:chiral-condensate}
    \hat{\mathcal{C}} = \frac{1}{L_xL_y} \sum_\mathbf{r} (-1)^{r_x+r_y} \left( \hat{\phi}_\mathbf{r}^\dagger \hat{\phi}_\mathbf{r} - \frac{1 - (-1)^{r_x+r_y}}{2} \right),
\end{equation}
in panels (b), we show the von Neumann entanglement entropy \(\mathcal{S}(t)\) measured using a bipartition formed by a slice along the circumference of the cylinder, and in panels (c), we show the fidelity with the initial state \(\mathcal{F}(t) = \lvert\braket{\psi(0)}{\psi(t)}\rvert^2\) for the finite system.

For the quenches with bosonic matter (Fig.~\ref{fig:quench-b-cp}), we can see the signatures of QMBS are qualitatively preserved, although less pronounced, as \(\alpha\) increases from zero (where the dynamics will correspond to the $1+1$D QLM) up to one (where we retrieve the fully fledged $2+1$D QLM).
The oscillatory behavior of the chiral condensate (Fig.~\ref{fig:quench-b-cp}(a)) and entanglement entropy (Fig.~\ref{fig:quench-b-cp}(b)) is clearly visible for all values of \(\alpha\) throughout the whole simulation, and the fidelity (Fig.~\ref{fig:quench-b-cp}(c)) shows revivals of a consistent, though weaker, magnitude.
However, in the fermionic matter quenches (Fig.~\ref{fig:quench-f-cp}), the signatures of QMBS clearly break down as \(\alpha\) is increased.
The oscillations in the chiral condensate (Fig.~\ref{fig:quench-f-cp}(a)) quickly decay in magnitude, the growth of the entanglement entropy (Fig.~\ref{fig:quench-f-cp}(b)) is significantly more rapid, and the revivals in fidelity (Fig.~\ref{fig:quench-f-cp}(c)) also swiftly decrease in magnitude.
These results show that the fate of QMBS in the $2+1$D QLM is strongly dependent on the particle statistics of the matter degrees of freedom, as the signs of scarring are preserved for bosonic, but not fermionic, matter. Nevertheless, it is quite an interesting finding to see that QMBS persists and is rather robust in $d=2$ spatial dimensions in the presence of hard-core bosonic dynamical matter.

We also find that the plaquette term weakens the QMBS observed in Figs.~\ref{fig:quench-b-cp} and~\ref{fig:quench-f-cp}, with more suppressed QMBS the larger $J$ is; see Supplemental Material (SM) \cite{SM}. We have also investigated quenching from the vacuum states of the $2+1$D QLM, but we have found no signs of QMBS regardless of whether the matter degrees of freedom are fermionic or hard-core bosonic \cite{SM}.

\begin{figure}[t]
    \includegraphics{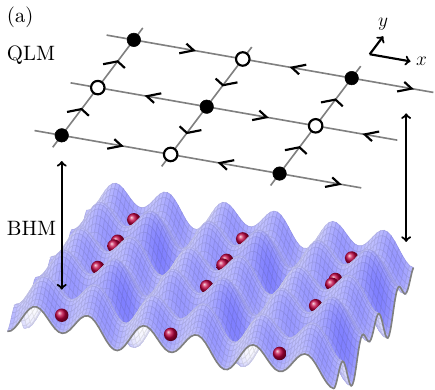}
    \includegraphics{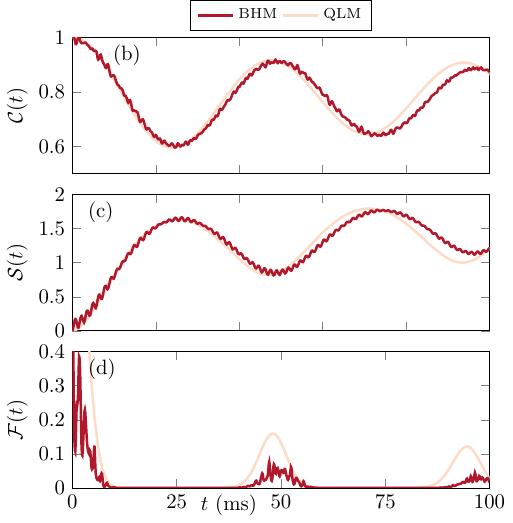}
    \caption{(a) The mapping between the $2+1$D \(\mathrm{U}(1)\) QLM and the Bose–Hubbard simulator proposed in Ref.~\cite{osborne2022largescale}.
    (b–d) A numerical time-evolution simulation of the quench of the charge-proliferated state shown in (a) to \(m = 0.84 \kappa\) on a \(L_y = 4\) cylinder for Bose--Hubbard simulator~\eqref{eq:bhm}, compared with the QLM simulation in Fig.~\ref{fig:quench-b-cp}, showing the chiral condensate \(\mathcal{C}(t)\)~\eqref{eq:chiral-condensate}, entanglement entropy \(\mathcal{S}(t)\), and fidelity with the initial state \(\mathcal{F}(t)\).}
    \label{fig:bhm}
\end{figure}

\textbf{\textit{Quantum simulation.---}}In order to demonstrate the relevance of these results to near-term quantum simulators, we show that the dynamics for the bosonic QLM (Fig.~\ref{fig:quench-b-cp}) can be probed using the 2D Bose–Hubbard simulator proposed in Ref.~\cite{osborne2022largescale} (see Fig.~\ref{fig:bhm}(a) for the mapping of the charge-proliferated state), with the Hamiltonian
\begin{multline}\label{eq:bhm}
    \hat{H}_\text{BHM} = \sum_\mathbf{j} \bigg[ \tilde{J} \sum_{a=x,y} \left( \hat{b}_\mathbf{j}^\dagger \hat{b}_{\mathbf{j}+\mathbf{e}_a} + \text{H.c.} \right) \\
    + \frac{U_\mathbf{j}}{2} \hat{n}_\mathbf{j} (\hat{n}_\mathbf{j}-1) + (\vec{\gamma} \cdot \mathbf{j} - \delta_\mathbf{j} - \eta_\mathbf{j}) \hat{n}_\mathbf{j} \bigg].
\end{multline}
Here, the lattice site of index \(\mathbf{j} = (j_x, j_y)\) corresponds to a matter site in the QLM if both components are even, to a gauge site if exactly one of them is even, or if both components are odd the site is forbidden from contributing to the dynamics.
\(\hat{b}_\mathbf{j}^{(\dagger)}\) and \(\hat{n}_\mathbf{j} = \hat{b}_\mathbf{j}^\dagger \hat{b}_\mathbf{j}\) are the bosonic ladder and number operators respectively, the on-site interaction strength \(U_\mathbf{j}\) is equal to \(U\) on gauge and forbidden sites, but \(\tilde{\alpha} U\) on matter sites, for some detuning constant \(\tilde{\alpha}\), \(\vec{\gamma} = (\gamma_x, \gamma_y)\) represents a linear tilt in the lattice in both dimensions, and the potential \(\delta_\mathbf{j}\) (\(\eta_\mathbf{j}\)) is equal to \(\delta\) (\(\eta\)) only on a gauge (forbidden) site, and zero elsewhere.
Following Ref.~\cite{osborne2022largescale}, we use the values of the parameters \(\tilde{J} = 30\,\text{Hz}\), \(U = 1300\,\text{Hz}\), \(\tilde{\alpha} = 1.3\), \(\gamma_x = 57\,\text{Hz}\), \(\gamma_y = 73\,\text{Hz}\), \(\delta = 649.647\,\text{Hz}\), and \(\eta = 5\delta\).
This corresponds to \(m \approx -0.84 \kappa\) in the QLM (quenches starting from the charge-proliferated initial state Fig.~\ref{fig:1}(b) have the same dynamics for positive and negative \(m\)).
The results of the quench of the charge-proliferated initial state in Fig.~\ref{fig:bhm}(a) are shown in Fig.~\ref{fig:bhm}(b–d), compared with the QLM simulation (Fig.~\ref{fig:quench-b-cp}) using the simulator’s effective value of \(\kappa\), which show great agreement for the available simulation times. This bodes well for upcoming cold-atom experiments seeking to realize $2+1$D $\mathrm{U}(1)$ QLMs, as it brings our findings into the realm of experimental accessibility.

\textbf{\textit{Discussion and outlook.---}}We have studied QMBS in a spin-$1/2$ $2+1$D $\mathrm{U}(1)$ QLM with dynamical matter using MPS simulations of quench dynamics. We have found that QMBS is robust when the initial state is the charge-proliferated product state and the quench mass is $m\approx0.84\kappa$, particularly when the matter degrees of freedom are hard-core bosonic. In the case of fermionic matter, we have found that scarring persists only for short times, but then vanishes and the dynamics looks quite ergodic. We have also adopted in our simulations a ratio $\alpha$ between coupling in the $x$ and $y$ directions. When $\alpha=0$, we are effectively in $1+1$D, and so we can retrieve the established QMBS regimes there. Upon tuning $\alpha$ up to unity (the fully fledged $2+1$D system), we see that scarring unsurprisingly gets weaker, but does not always vanish, as explained above. In this work we have used $4\times \infty$ and $4\times 16$ cylinders, but we expect that the same qualitative picture remains for wider cylinders \cite{Hashizume2022dynamical,Hashizume2022}.

We have also showcased how our findings can be tested on a recently proposed cold-atom quantum simulator of the spin-$1/2$ $\mathrm{U}(1)$ QLM with hard-core bosonic matter degrees of freedom \cite{osborne2022largescale}, showing great agreement between the simulator dynamics and those of the ideal QLM. But our results are also amenable for investigation in other proposals that have been put forth \cite{Zohar2013,Ott2020scalable,Fontana2022,surace2023abinitio}. Given that going to $d=2$ spatial dimensions in quantum-simulator realizations of LGTs is the current frontier of the field \cite{Zohar_NewReview}, our work sets the stage for experimentally relevant features that can be probed on such devices once they are available.

It is important to emphasize that our work does not rule out other QMBS regimes that may still persist in $2+1$D but that we have not found. Indeed, our goal in this work was to investigate the fate of QMBS regimes discovered in the $1+1$D $\mathrm{U}(1)$ QLM and see how well they fare in $2+1$D. It was also our intention to find out how the statistics of the matter degrees of freedom affects QMBS, which, as we show, happens in a profound way. One avenue for future work is to study the fate of the QMBS regime we find in $2+1$D for higher-level representations ($S>1/2$) of the electric and gauge fields. In $1+1$D, it has been shown that scarring persists and is robust for $S>1/2$ \cite{Desaules2022weak,Desaules2022prominent}. This is important to assess the fate of QMBS in the limit of $2+1$D lattice QED, and a recent proposal may allow experimental observation in a quantum simulator of the spin-$1$ $\mathrm{U}(1)$ QLM \cite{osborne2023spins}. A second venue involves adding a topological $\theta$-term \cite{Halimeh2022tuning,Cheng2022tunable} and studying its interplay with scarring in $2+1$D, which has been shown to lead to a plethora of nonergodic behavior is $1+1$D \cite{Desaules2024ergodicitybreaking}.

\begin{acknowledgments}
\textbf{\textit{Note.---}}During the final stages of preparing our manuscript, we became aware of another work \cite{Budde2024} on quantum many-body scars for arbitrary integer spin in $2+1$D pure Abelian gauge theories. This work will appear in the same \texttt{arXiv} listing as ours.

\textbf{\textit{Acknowledgments.---}}The authors are grateful to Jean-Yves Desaules and Zlatko Papi\'c for fruitful discussions. This work is supported by the Emmy Noether Programme of the German Research Foundation (DFG) under grant no.~HA 8206/1-1. I.P.M.~acknowledges funding from the National Science and Technology Council (NSTC) Grant No.~122-2811-M-007-044.
Numerical simulations were performed on The University of Queensland's School of Mathematics and Physics Core Computing Facility \texttt{getafix}.
\end{acknowledgments}

\bibliography{biblio}

\clearpage
\pagebreak
\newpage
\appendix
\setcounter{equation}{0}
\setcounter{figure}{0}
\setcounter{table}{0}
\setcounter{page}{1}
\makeatletter
\renewcommand{\theequation}{S\arabic{equation}}
\renewcommand{\thefigure}{S\arabic{figure}}
\renewcommand{\thetable}{S\Roman{table}}
\widetext
\begin{center}
\textbf{--- Supplemental Material ---\\Quantum Many-Body Scarring in $2+1$D Gauge Theories with Dynamical Matter}
\text{Jesse Osborne, Ian P.~McCulloch, and Jad C.~Halimeh}
\end{center}

\begin{figure}[h]
    \begin{minipage}{0.48\linewidth}
        \includegraphics{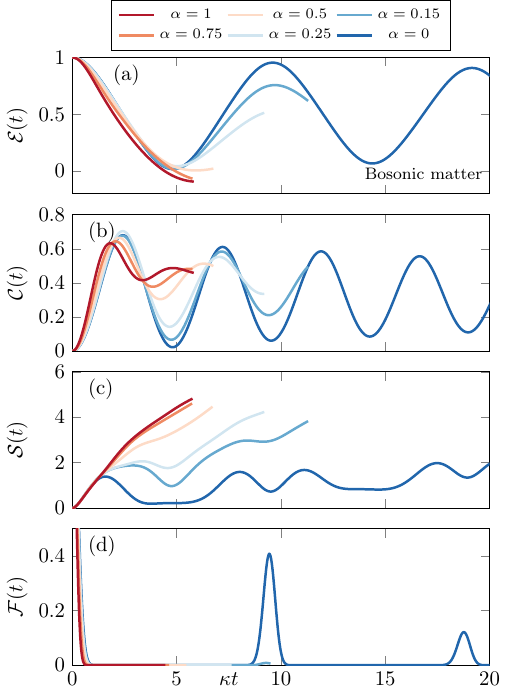}
        \caption{A numerical time-evolution simulation of the quench of the vacuum state shown in Fig.~\ref{fig:1}(c) to \(m = 0\), using matter with bosonic statistics on a \(L_y = 4\) cylinder, as the coupling ratio \(\alpha\) tuned from 1 to 0.
        (a) The flux \(\mathcal{E}(t)\)~\eqref{eq:flux}, (b) chiral condensate \(\mathcal{C}(t)\)~\eqref{eq:chiral-condensate}, and (c) entanglement entropy across a circumferential slice \(\mathcal{S}(t)\), calculated for an infinite-length cylinder.
        (d) The fidelity \(\mathcal{F}(t)\), calculated for a finite cylinder of dimension \(4 \times 16\) with open boundaries in the \(x\) direction.}
        \label{fig:quench-b-v}
    \end{minipage}\hfill
    \begin{minipage}{0.48\linewidth}
        \includegraphics{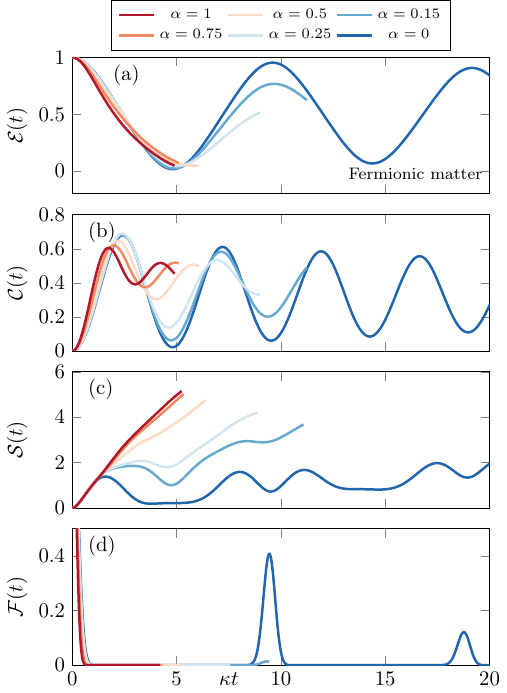}
        \caption{A numerical time-evolution simulation of the quench of the vacuum state shown in Fig.~\ref{fig:1}(c) to \(m = 0\), using matter with fermionic statistics on a \(L_y = 4\) cylinder, as the coupling ratio \(\alpha\) tuned from 1 to 0.
        (a) The flux \(\mathcal{E}(t)\)~\eqref{eq:flux}, (b) chiral condensate \(\mathcal{C}(t)\)~\eqref{eq:chiral-condensate}, and (c) entanglement entropy across a circumferential slice \(\mathcal{S}(t)\), calculated for an infinite-length cylinder.
        (d) The fidelity \(\mathcal{F}(t)\), calculated for a finite cylinder of dimension \(4 \times 16\) with open boundaries in the \(x\) direction.}
        \label{fig:quench-f-v}
    \end{minipage}
\end{figure}

\section{Quenches of the vacuum state}
In Figures~\ref{fig:quench-b-v} and \ref{fig:quench-f-v}, we show quenches of a vacuum initial state (Fig.~\ref{fig:1}(c) in the main text) to \(m = 0\).
Here, we also show the expectation value of the electric flux \(\mathcal{E}(t) = \mel{\psi(t)}{\hat{\mathcal{E}}}{\psi(t)}\),
\begin{equation}\label{eq:flux}
    \hat{\mathcal{E}} = \frac{1}{L_xL_y} \sum_\mathbf{r} \sum_{a=x,y} \hat{s}^z_{\mathbf{r},\mathbf{e}_a}.
\end{equation}
In the $1+1$D case (\(\alpha = 0\)), this quench exhibits QMBS, but for \(\alpha > 0\), the signs of QMBS are suppressed for both bosonic (Fig.~\ref{fig:quench-b-v}) and fermionic (Fig.~\ref{fig:quench-f-v}) matter fields, with little qualitative difference between the two different particle statistics.

\begin{figure}[h]
    \begin{minipage}{0.48\linewidth}
        \includegraphics{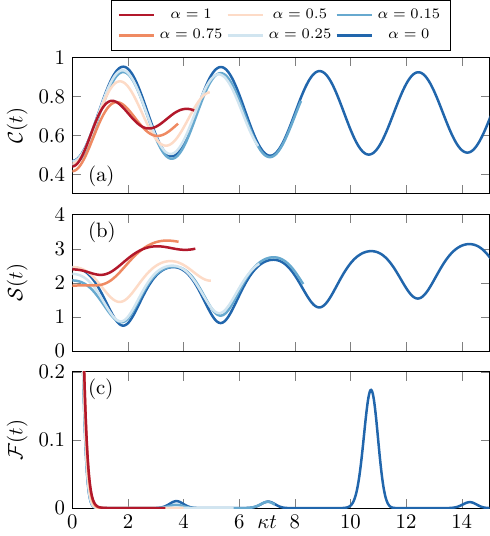}
        \caption{A numerical time-evolution simulation of the quench of a gauge-invariant ground state at \(m_i = 0.19\kappa\) to \(m_f = -0.4\kappa\), using matter with bosonic statistics on a \(L_y = 4\) cylinder, as the coupling ratio \(\alpha\) tuned from 1 to 0.
        (a) The chiral condensate \(\mathcal{C}(t)\)~\eqref{eq:chiral-condensate} and (b) entanglement entropy across a circumferential slice \(\mathcal{S}(t)\), calculated for an infinite-length cylinder.
        (c) The fidelity \(\mathcal{F}(t)\), calculated for a finite cylinder of dimension \(4 \times 16\) with open boundaries in the \(x\) direction.}
        \label{fig:quench-b-finite-m}
    \end{minipage}\hfill
    \begin{minipage}{0.48\linewidth}
        \includegraphics{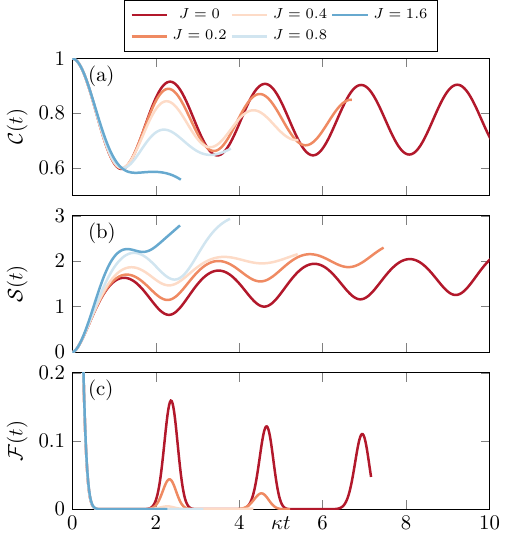}
        \caption{A numerical time-evolution simulation of the quench of the charge-proliferated state shown in Fig.~\ref{fig:1}(b) to \(m = 0.84 \kappa\), using matter with bosonic statistics on a \(L_y = 4\) cylinder, as the magnetic coupling \(J\) is increased, fixing the coupling ratio \(\alpha\) to be 1.
        (a) The chiral condensate \(\mathcal{C}(t)\)~\eqref{eq:chiral-condensate} and (b) entanglement entropy across a circumferential slice \(\mathcal{S}(t)\), calculated for an infinite-length cylinder.
        (c) The fidelity \(\mathcal{F}(t)\), calculated for a finite cylinder of dimension \(4 \times 16\) with open boundaries in the \(x\) direction.}
        \label{fig:quench-b-cp-J}
    \end{minipage}
\end{figure}

\section{Quenches starting at a finite mass}
Here, we examine a sudden quench from a finite value of the mass, starting with the gauge-invariant ground state at \(m_i = 0.19\kappa\), and evolving the state at \(m_f = -0.4\kappa\), which was shown to display strong signs of QMBS in the $1+1$D model in Ref.~\cite{Su2022}.
As shown in Figure~\ref{fig:quench-b-finite-m}, however, for \(\alpha > 0\) these signs are strongly suppressed.
The way the gauge-invariant ground state at \(m_i = 0.19\kappa\) changes as \(\alpha\) is increased from zero to one would play a significant role here, and requires further investigation.

\section{Effect of the magnetic coupling of the gauge fields}
In Figure~\ref{fig:quench-b-cp-J}, we examine the effect of the magnetic coupling of the gauge fields, controlled by the plaquette term with coefficient \(J\) in the \(2+1\)D QLM Hamiltonian, Eq.~\eqref{eq:qlm} in the main text.
We look at the quench from the charge proliferated state (Fig.~\ref{fig:1}(b)) to \(m = 0.84\kappa\) for bosonic matter, which was shown to display strong signs of QMBS in the main text (Fig.~\ref{fig:quench-b-cp}).
However, for \(J > 0\), we can see that these signs are quickly suppressed.
\end{document}